\documentclass[english,aip,pop,reprint,floatfix]{revtex4-2}
\usepackage[T1]{fontenc}
\usepackage[latin9]{inputenc}
\usepackage{amsmath}
\usepackage{graphicx}
\usepackage[normalem]{ulem} 
\newcommand{\replace}[2]{#2}
\usepackage{babel}
\usepackage{color}
\usepackage{url}
\begin{document}
\title{Computation of Magnetohydrodynamic Equilibria with Voigt Regularization }
\author{Yi-Min Huang}
\affiliation{Department of Astronomy, University of Maryland, College Park, Maryland
20742, USA}
\email{yopology@umd.edu}

\affiliation{National Aeronautics and Space Administration Goddard Space Flight
Center, Greenbelt, Maryland 20771, USA}
\author{Justin Kin Jun Hew}
\affiliation{ACCESS-NRI, Australian National University, Canberra, ACT 2601, Australia}
\affiliation{Space Plasma Power and Propulsion Laboratory, Research School of Physics,
Australian National University, Canberra, ACT 2601, Australia}
\author{Andrew Brown}
\affiliation{Department of Astrophysical Sciences, Princeton University, New Jersey 08543, USA}
\author{Amitava Bhattacharjee}
\affiliation{Department of Astrophysical Sciences, Princeton University, New Jersey
08543, USA}
\begin{abstract}
This work presents the first numerical investigation of using Voigt regularization
as a method for obtaining magnetohydrodynamic (MHD) equilibria
without the assumption of nested magnetic flux surfaces. Voigt regularization
modifies the MHD dynamics by introducing additional terms that vanish
in the infinite-time limit, allowing for magnetic reconnection and
the formation of magnetic islands, which can overlap and produce field-line chaos.
The utility of this approach is demonstrated through numerical solutions
of two-dimensional ideal and resistive test problems. Our results
show that Voigt regularization can significantly accelerate the convergence
to solutions in resistive MHD problems, while also highlighting challenges
in applying the method to ideal MHD systems. This research opens up
new possibilities for developing more efficient and robust MHD equilibrium
solvers, which could contribute to the design and optimization of
future fusion devices.
\end{abstract}
\maketitle

\section{Introduction}

Magnetohydrodynamic (MHD) equilibrium solutions, with and without flows, have a wide range
of applications in space and laboratory plasma experiments, including
planetary magnetospheres, stellar atmospheres,\citep{Priest2014} and
magnetic fusion devices.\citep{Freidberg1987} For magnetic confinement fusion applications, equilibria without flow (also referred to as magnetohydrostatic (MHS)) are usually a first point of departure. Furthermore, solutions with a continuum of nested flux surfaces are desirable for fusion devices.
Therefore, many three-dimensional (3D) equilibrium solvers assume this as a starting point.\citep{HirshmanW1983,Garabedian2002,DudtK2020}
It is well-known, however, that ideal three-dimensional (3D) toroidal MHS equilibria with nested flux surfaces are usually ill-behaved at rational surfaces where magnetic field lines close on themselves.\citep{Grad1967} Specifically, delta-function and algebraically diverging singularities of current density can arise at those surfaces.\citep{bhattacharjee1995,helander2014} These singular current densities are integrable and do not pose a problem as far as weak
solutions are concerned.\citep{HuangHLZB2022,HuangZLHB2023} Their
presence, however, indicates that those flux surfaces in a real plasma
will tear and reconnect, forming magnetic islands, which can overlap and produce chaotic field-line regions. Moreover, because rational surfaces are dense if the
rotational transform $\iota$ has a continuous profile, the corresponding
ideal MHS equilibrium may have densely distributed singularities throughout
the entire volume. Calculating numerical approximations for such an
equilibrium is a formidable challenge. 

To resolve this issue of densely distributed singularities in 3D equilibria, approaches to regularize MHS equilibria have been proposed. A prominent
example is the multi-region relaxed magnetohydrodynamics (MRxMHD).\citep{HoleHD2007}
In the MRxMHD description, the entire plasma volume is divided into
multiple regions. In each region, the magnetic field relaxes to a
Beltrami field (also known as a Taylor state) satisfying $\nabla\times\boldsymbol{B}=\mu\boldsymbol{B}$
while conserving magnetic helicity in that domain. The
force-balance condition is enforced across the interfaces between
regions. Because Beltrami fields are force-free, the plasma pressure
in each region is constant. Over the entire volume, the plasma pressure
comprises a sequence of step functions, with discontinuities at the interfaces
between regions. To maintain force-balance across the interfaces with
discontinuous pressure, the magnetic field must also be discontinuous,
corresponding to delta-function singular current densities. The MRxMHD
model bridges Taylor relaxation\citep{Taylor1974}
and ideal MHD: When the entire volume contains only one region, MRxMHD
is equivalent to Taylor relaxation, whereas in the limit of infinite
number of regions, MRxMHD approaches ideal MHD under certain conditions.\citep{DennisHDH2013,QuHDLH2021}
A numerical code to solve MRxMHD equilibria, the Stepped Pressure
Equilibrium Code (SPEC), has been developed by Hudson \emph{et al.}\citep{HudsonDDHMNL2012}

One of the strong attributes of MRxMHD is that it allows magnetic islands and field-line chaos within each region,
but the interfaces between regions form a set of nested surfaces that
cannot break. The placement of the interfaces is arbitrary and remains
a critical issue of MRxMHD research.\citep{LoizuHHBKQ2020,QuHDLH2021,BalkovicLGHS2024} Usually the interfaces are placed at surfaces of strongly irrational rotational
transform $\iota$ (i.e., it satisfies the Diophantine condition)
because those flux surfaces are more difficult to break.\citep{KrausH2017}
More generally, MRxMHD also permits solutions with discontinuous rotational
transform profiles.\citep{LoizuHBLH2015} 

The uncertainties in the choice of interfaces can pose challenges
in constructing MRxMHD solutions, motivating consideration of other approaches to regularizing the MHD equilibrium. For example, Dewar and
Qu have recently proposed an augmented Lagrangian multiplier method
to approximate the ideal MHD Ohm's law.\citep{DewarQ2022} This
method avoids the exact enforcement of the ideal Ohm's law constraint,
which is the source of current singularities. The present study
considers another approach to regularizing the MHD equilibrium using the so-called Voigt approximation. 

This approach is motivated by recent mathematical theorems of Constantin and Pasqualotto.\citep{ConstantinP2023} The theorems assert that the infinite-time limit of the Voigt approximations of viscous, non-resistive, incompressible MHD equations are MHS equilibria that are regular {(in a specific Sobolev space)} and non-Beltrami. {This result suggests that MHS solutions could be obtained by numerically solving the Voigt-MHD system forward in time until it reaches an equilibrium.} The Voigt approximation modifies MHD dynamics by introducing additional terms {involving higher-order spatial operators acting on the time derivatives of field variables in} the governing equations. These additional terms vanish as the solution approaches a steady state in the infinite-time limit; {therefore, they do not affect the MHD force balance. The additional term in the induction equation is analogous to the electron inertia effect, which allows magnetic reconnection and the formation of magnetic islands and chaotic field-line regions. We test the practicality of using Voigt-MHD as an equilibrium solver by implementing a two-dimensional (2D) version and applying it to ideal and resistive test problems. However, despite the theorems of Constantin and Pasqualotto that guarantee regular solutions, our ideal tests encounter difficulties due to the formation of very thin current sheets that require high numerical resolutions. In contrast, our tests of resistive problems yield positive results of rapid convergence to equilibria, due to the large time-steps enabled by the Voigt regularization.}

{The approach of Voigt-MHD bears some similarities with existing MHS solvers that do not assume nested flux surfaces. For example, the HINT2 code alternately relaxes the pressure and the magnetic field in a way that mimics time evolution.\citep{SuzukiNWNH2006, Suzuki2017} In the pressure relaxation step, the new pressure is obtained at each point by averaging the current pressure along the field line over a given length. The new pressure is then used in the magnetic field relaxation step, which solves the momentum and induction equations of the resistive MHD system, holding the pressure fixed. These steps continue until an equilibrium is reached. In an alternative approach, the SIESTA code directly steps the adiabatic pressure equation and the induction equation simultaneously.\cite{HirshmanSC2011} However, a small resistivity has to be periodically applied to allow breaking of magnetic surfaces. The Voigt-MHD approach differs from these methods by being grounded in a well-defined system of partial differential equations rather than a prescription of somewhat ad hoc procedures.}

This paper is organized as follows. Section \ref{sec:Voigt-MHD} lays out the formalism of the Voigt-MHD equations and derives the energy conservation
law and the wave dispersion relation. The Voigt approximation modifies the
form of total energy, which involves current density and vorticity. This modified energy form provides critical insight into why the Voigt approximation regularizes the solution. The wave dispersion relation further suggests that the Voigt approximation slows down MHD waves, allowing larger time-steps and potentially expediting the convergence to equilibrium. Section
\ref{sec:Two-dimensional-Test-Problems} assesses the feasibility
and effectiveness of using the Voigt approximation by numerically solving
two-dimensional (2D) test problems of MHD equilibria. We first describe a 2D pseudospectral numerical solver of Voigt-MHD in slab geometry using the Dedalus framework.\citep{BurnsVOLB2020} We then apply the solver to a few ideal and resistive test problems, including the saturation
of the tearing instability\citep{LoizuHHBKQ2020} and the Hahm--Kulsrud--Taylor forced reconnection problem.\citep{HahmK1985} These test problems have a common theme of magnetic island formation. Section \ref{sec:Conclusions-and-Future}
discusses the implications of our findings, open questions, and future
perspectives of applying the Voigt regularization for the development
of more robust and efficient MHD equilibrium solvers.

\section{Voigt Regularization of Incompressible Magnetohydrodynamics\protect\label{sec:Voigt-MHD}}

\subsection{Governing Equations and Boundary Conditions}

The non-dimensional form of the Voigt-regularized, incompressible MHD
system is given by
\begin{equation}
\partial_{t}\left(\boldsymbol{u}-\alpha_{1}\nabla^{2}\boldsymbol{u}\right)+\boldsymbol{u}\cdot\nabla\boldsymbol{u}=-\nabla p+\boldsymbol{J}\times\boldsymbol{B}+\nu\nabla^{2}\boldsymbol{u},\label{eq:momentum}
\end{equation}
\begin{equation}
\partial_{t}\left(\boldsymbol{B}-\alpha_{2}\nabla^{2}\boldsymbol{B}\right)=\nabla\times\left(\boldsymbol{u}\times\boldsymbol{B}-\eta\boldsymbol{J}-\boldsymbol{E}_{\text{ext}}\right),\label{eq:induction}
\end{equation}
\begin{equation}
\boldsymbol{J}=\nabla\times\boldsymbol{B},\label{eq:J_from_B}
\end{equation}
\begin{equation}
\nabla\cdot\boldsymbol{u}=\nabla\cdot\boldsymbol{B}=0.\label{eq:constraints}
\end{equation}
Standard notations are used: $\boldsymbol{u}$ is the plasma velocity,
$\boldsymbol{B}$ is the magnetic field, $p$ is the pressure, $\boldsymbol{J}$
is the current density, $\nu$ is the viscosity coefficient, and $\eta$
is the resistivity. Voigt regularization adds two terms $-\alpha_{1}\partial_t\nabla^{2}\boldsymbol{u}$ and $-\alpha_{2}\partial_t\nabla^{2}\boldsymbol{B}$,
where the coefficients $\alpha_{1}$ and $\alpha_{2}$ are free parameters.

The induction equation (\ref{eq:induction}) corresponds to a generalized
Ohm's law 
\begin{equation}
\boldsymbol{E}=-\boldsymbol{u}\times\boldsymbol{B}+\alpha_{2}\partial_{t}\boldsymbol{J}+\eta\boldsymbol{J}+\boldsymbol{E}_{\text{ext}}.\label{eq:Ohm}
\end{equation}
When $\eta>0$ , an external electric field $\boldsymbol{E}_{\text{ext}}$
is applied to prevent the solution from decaying to a vacuum field.
The term $\alpha_{2}\partial_{t}\boldsymbol{J}$ coming from Voigt
regularization is analogous to the electron inertia term in a generalized Ohm's law.\citep{GurnettB2017}
Similar to the electron inertia effect, the $\alpha_{2}\partial_{t}\boldsymbol{J}$
term breaks the frozen-in condition, allowing magnetic reconnection
even when the resistivity vanishes.\cite{GrassoPPC1999}

The boundaries are assumed to be impenetrable, no-slip, and perfectly
conducting, leading to the following boundary conditions: 
\begin{equation}
\left.\boldsymbol{u}\right|_{\text{wall}}=0,\label{eq:BC}
\end{equation}
 and
\begin{equation}
\left.\boldsymbol{\hat{n}}\times\boldsymbol{E}\right|_{\text{wall}}=\left.\boldsymbol{\hat{n}}\times\left[\alpha_{2}\partial_{t}\boldsymbol{J}+\eta\boldsymbol{J}+\boldsymbol{E}_{\text{ext}}\right]\right|_{\text{wall}}=0,\label{eq:BC1}
\end{equation}
where $\boldsymbol{\hat{n}}$ is the unit normal vector to the wall. {The perfectly conducting boundary condition implies $\left.\boldsymbol{\hat{n}}\cdot\partial_t \boldsymbol{B}\right|_{\text{wall}}=0$.} 

\replace{}{This set of equations does not include an explicit equation for the pressure $p$. The pressure is obtained through the incompressible constraint $\nabla \cdot \boldsymbol{u}=0$, which implies [by taking the divergence of Eq.~(\ref{eq:momentum})]
\begin{equation}
    \nabla^2 p = \nabla\cdot\left(-\boldsymbol{u}\cdot\nabla\boldsymbol{u}+\boldsymbol{J}\times\boldsymbol{B} \right).
    \label{eq:poisson}
\end{equation}
This Poisson equation must be supplied with an appropriate boundary condition such that the velocity, when stepped with Eq.~(\ref{eq:momentum}), satisfies the boundary condition (\ref{eq:BC})}.

An important feature of the Voigt regularization is that all the additional
terms are within the time derivative terms of the momentum and induction equations. In the infinite-time limit
when the solution approaches a steady state, those terms vanish. Therefore,
the Voigt-regularization terms have no effect on the force-balance
once a steady-state solution is obtained. 

\subsection{Energy Conservation}

The energy conservation law for the Voigt-MHD system can be obtained
by adding the inner product of Eq. (\ref{eq:momentum}) with $\boldsymbol{u}$
and the inner product of Eq. (\ref{eq:induction}) with $\boldsymbol{B}$,
and integrating the sum over the entire volume. We break
the derivation into several parts. First, note that for an incompressible
flow, the condition $\nabla^{2}\boldsymbol{u}=-\nabla\times\boldsymbol{\omega}$
is satisfied, where $\boldsymbol{\omega}=\nabla\times\boldsymbol{u}$
is the vorticity. We can then derive the equation 
\begin{equation}
\boldsymbol{u}\cdot\nabla^{2}\boldsymbol{u}=-\boldsymbol{u}\cdot\nabla\times\boldsymbol{\omega}=\nabla\cdot\left(\boldsymbol{u}\times\boldsymbol{\omega}\right)-\omega^{2},\label{eq:u_dot_del2_u}
\end{equation}
by using vector identities. Likewise, we have
\begin{equation}
\boldsymbol{B}\cdot\nabla^{2}\boldsymbol{B}=\nabla\cdot\left(\boldsymbol{B}\times\boldsymbol{J}\right)-J^{2}.\label{eq:B_dot_del2_B}
\end{equation}
From Eqs. (\ref{eq:u_dot_del2_u}) and (\ref{eq:B_dot_del2_B}), we
obtain
\begin{equation}
\boldsymbol{u}\cdot\partial_{t}\nabla^{2}\boldsymbol{u}=\nabla\cdot\left(\boldsymbol{u}\times\partial_{t}\boldsymbol{\omega}\right)-\partial_{t}\frac{\omega^{2}}{2},\label{eq:dt1}
\end{equation}
and 
\begin{equation}
\boldsymbol{B}\cdot\partial_{t}\nabla^{2}\boldsymbol{B}=\nabla\cdot\left(\boldsymbol{B}\times\partial_{t}\boldsymbol{J}\right)-\partial_{t}\frac{J^{2}}{2}.\label{eq:dt2}
\end{equation}

Next, the identities 
\begin{equation}
\boldsymbol{B}\cdot\nabla\times\left(\boldsymbol{u}\times\boldsymbol{B}\right)=\nabla\cdot\left(\left(\boldsymbol{u}\times\boldsymbol{B}\right)\times\boldsymbol{B}\right)-\boldsymbol{u}\cdot\left(\boldsymbol{J}\times\boldsymbol{B}\right),\label{eq:identity1}
\end{equation}
and 
\begin{multline}
\boldsymbol{B}\cdot\nabla\times\left(\eta\boldsymbol{J}+\boldsymbol{E}_{\text{ext}}\right)\\
=\nabla\cdot\left(\left(\eta\boldsymbol{J}+\boldsymbol{E}_{\text{ext}}\right)\times\boldsymbol{B}\right)+\eta J^{2}+\boldsymbol{E}_{\text{ext}}\cdot\boldsymbol{J}\label{eq:identity2}
\end{multline}
are readily derived.

Using Eqs (\ref{eq:u_dot_del2_u}) -- (\ref{eq:identity2}), the
energy conservation law can be written as 
\begin{align}
 & \frac{d}{dt}\int d^{3}x\left(\frac{u^{2}+\alpha_{1}\omega^{2}+B^{2}+\alpha_{2}J^{2}}{2}\right)\nonumber \\
= & -\int d^{3}x\left(\nu\omega^{2}+\eta J^{2}\right)-\int d^{3}x\boldsymbol{E}_{\text{ext}}\cdot\boldsymbol{J},\label{eq:energy}
\end{align}
where surface integral terms {involving $\boldsymbol{u}$ vanish due to the boundary condition (\ref{eq:BC}) and the surface integral terms $-\int_S dS\boldsymbol{\hat{n}}\cdot\left(\alpha_{2}\partial_{t}\boldsymbol{J}\times\boldsymbol{B}\right)-\int_S dS\boldsymbol{\hat{n}}\cdot\left(\left(\eta\boldsymbol{J}+\boldsymbol{E}_{ext}\right)\times\boldsymbol{B}\right)$ also vanish because of the boundary condition (\ref{eq:BC1}).} The left-hand-side
of Eq.~(\ref{eq:energy}) is the time derivative of the ``energy''
in Voigt-MHD. The first term on the right-hand-side represents the
energy dissipated due to viscosity and resistivity, whereas the second
term accounts for the energy injected or extracted by the external
electric field. \replace{}{Note that the energy functional does not involve the pressure $p$, because the incompressible constraint is equivalent to the limit of the heat capacity ratio $\gamma\to\infty$; therefore, the thermal energy density $p/\left(\gamma-1\right)\to0$.}

The presence of $\alpha_{1}\omega^{2}+\alpha_{2}J^{2}$ in the energy functional provides a critical insight as to why the Voigt terms regularize the solution. For the ``ideal'' case with $\eta=0$ and $\boldsymbol{E}_{\text{ext}}=0$, the total energy remains bounded, which precludes the formation
of delta-function current density and vorticity, otherwise \replace{}{the integral of} $J^{2}$
or $\omega^{2}$ will {diverge}. Although energy consideration
alone cannot preclude algebraically divergent current density and
vorticity completely, their formation is strongly restricted by the
requirement of energy being bounded. More generally, Voigt terms of
the form $\alpha(-\nabla^2)^n$ can be considered if higher-order
regularity is desired.\citep{ConstantinP2023}

\subsection{Wave Dispersion Relation}

The wave dispersion relation provides insight into the behavior of
waves, which is critical for developing efficient and stable numerical
methods for solving the system. We consider the dispersion relation
of a plane wave in a uniform magnetic field. Let the background magnetic
field $\boldsymbol{B}=B\boldsymbol{\hat{z}}$, and the wave-number
$\boldsymbol{k}=k_{\parallel}\boldsymbol{\hat{z}}+k_{\perp}\boldsymbol{\hat{x}}$.
Assuming a plane wave of the form $\sim e^{i\boldsymbol{k}\cdot\boldsymbol{x}-i\omega t}$
and denoting perturbed quantities with tildes, linearizing Voigt-MHD
yields
\begin{equation}
-i\omega\left(1+k^{2}\alpha_{1}\right)\tilde{\boldsymbol{u}}=-i\boldsymbol{k}\tilde{p}-i\boldsymbol{k}B\tilde{B_{z}}+ik_{\parallel}B\tilde{\boldsymbol{B}}-\nu k^{2}\tilde{\boldsymbol{u}},\label{eq:linear1}
\end{equation}
\begin{equation}
-i\omega\left(1+k^{2}\alpha_{2}\right)\tilde{\boldsymbol{B}}=ik_{\parallel}B\tilde{\boldsymbol{u}}-\eta k^{2}\tilde{\boldsymbol{B}},\label{eq:linear2}
\end{equation}
and
\begin{equation}
\boldsymbol{k}\cdot\tilde{\boldsymbol{u}}=\boldsymbol{k}\cdot\tilde{\boldsymbol{B}}=0.\label{eq:linear3}
\end{equation}
Taking the inner product of Eq. (\ref{eq:linear1}) with $\boldsymbol{k}$
and using Eq. (\ref{eq:linear3}) yields the relation $\tilde{p}+B\tilde{B_{z}}=0$.
Therefore, Eq. (\ref{eq:linear1}) is simplified to 
\begin{equation}
-i\omega\left(1+k^{2}\alpha_{1}\right)\tilde{\boldsymbol{u}}=ik_{\parallel}B\tilde{\boldsymbol{B}}-\nu k^{2}\tilde{\boldsymbol{u}}.\label{eq:linear1a}
\end{equation}
From Eqs.~(\ref{eq:linear2}) and (\ref{eq:linear1a}), the dispersion
relation is the solution of 
\begin{equation}
\left[\omega\left(1+k^{2}\alpha_{2}\right)+i\eta k^{2}\right]\left[\omega\left(1+k^{2}\alpha_{1}\right)+i\nu k^{2}\right]=k_{\parallel}^{2}B^{2}.\label{eq:dispersion}
\end{equation}
In the ideal limit with $\eta=\nu=0$, the dispersion relation is
\begin{equation}
\omega_{\text{ideal}}=\frac{\pm k_{\parallel}B}{\sqrt{\left(1+k^{2}\alpha_{1}\right)\left(1+k^{2}\alpha_{2}\right)}}.\label{eq:ideal_dispersion}
\end{equation}

The dispersion relation shows that the Voigt term $-\alpha_{1}\nabla^{2}\boldsymbol{u}$ in the momentum equation acts as a wave-number-dependent inertia
that slows down the frequencies of high-$k$ modes. \replace{}{The $-\alpha_1 \nabla^2$ operator is analogous to semi-implicit operators employed in several MHD codes to improve numerical stability when using large time-steps.\cite{SchnackBMHC1987, SovinecGGBNKSPTCN2004, Jardin2012}} Likewise, the term
$-\alpha_{2}\nabla^{2}\boldsymbol{B}$, analogous to the electron inertia effect, in the induction equation also slows down the frequencies of high-$k$ modes. This wave-slowing feature suggests that Voigt-regularization can
mitigate the {method of lines stability condition. The condition requires the eigenvalues of the linearized spatial discretization operator, scaled by the time-step $\Delta t$ to be within the stability
region of the time-discretization operator.\citep{Trefethen2000} By decreasing the wave frequencies, the Voigt regularization may allow a larger $\Delta t$,}
potentially reducing the computational cost and accelerating the convergence
to equilibrium solutions.

\section{Two-dimensional Test Problems\protect\label{sec:Two-dimensional-Test-Problems}}

Now we assess the feasibility of using Voigt regularization to obtain
MHD equilibria by solving 2D test problems in slab geometry. Section
\ref{subsec:Dedalus-Implementation} describes the implementation
of 2D Voigt-MHD using the Dedalus framework. Section \ref{subsec:Ideal-MHD-Equilibria}
applies the code to ideal MHD equilibria, and Section \ref{subsec:Resistive-MHD-Equilibria}
to resistive MHD equilibria with an external electric field.

\subsection{Dedalus Implementation of 2D Voigt-MHD \protect\label{subsec:Dedalus-Implementation}}

In a Cartesian coordinate system $(x,y,z)$, we assume that $z$ is
the direction of symmetry. We also assume that the $z$ components
of the magnetic field and the flow velocity both vanish. In this 2D
system, we can express the magnetic field in terms of the flux function
$\psi$ as 
\begin{equation}
\boldsymbol{B}=\nabla\psi\times\boldsymbol{\hat{z}}.\label{eq:B}
\end{equation}
The only non-vanishing component of the current density is along the
$z$ direction:
\begin{equation}
J=(\nabla\times\boldsymbol{B})_{z}=-\nabla^{2}\psi.\label{eq:J}
\end{equation}
With these definitions, the remaining equations are the momentum equation
\begin{equation}
\partial_{t}\left(\boldsymbol{u}-\alpha_{1}\nabla^{2}\boldsymbol{u}\right)+\boldsymbol{u}\cdot\nabla\boldsymbol{u}=-\nabla p-JB_{y}\boldsymbol{\hat{x}}+JB_{x}\boldsymbol{\hat{y}}+\nu\nabla^{2}\boldsymbol{u},\label{eq:momentum2d}
\end{equation}
and the induction equation
\begin{equation}
\partial_{t}\left(\psi-\alpha_{2}\nabla^{2}\psi\right)=\boldsymbol{u}\times\boldsymbol{B}-\eta J-E_{\text{ext}}.\label{eq:induction2d}
\end{equation}
In addition, the incompressible constraint $\nabla\cdot\boldsymbol{u} = 0$
is imposed.

This system of equations is numerically implemented in a rectangular
domain with a Fourier-Chebyshev pseudospectral method using the Dedalus
framework (\url{https://dedalus-project.org/}). The $x$ direction is assumed to be periodic, using a Fourier
representation truncated at $N_{x}$ modes. The $y$ direction is
bounded by perfectly conducting walls. The $y$ direction is discretized
with a composite of multiple Chebyshev segments. For the studies reported
in this paper, we use three segments along the $y$ direction, with each
segment employing $N_{y}$ Chebyshev polynomials. The middle segment
is narrower to provide a higher resolution near the mid-plane, where
a thin current sheet tends to form in the test problems. Specifically,
the domain along the $y$ direction is within the region $\left[-L_{y}/2,L_{y}/2\right]$
and the middle segment corresponds to the region $\left[-L_{y}/20,L_{y}/20\right]$.
The domain along the $x$ direction is within the range $\left[-L_{x}/2,L_{x}/2\right]$.
A dealiasing factor of $3/2$ is applied in both directions. Dedalus
has several options for time-stepping schemes. In this study, we use
the third-order, four-stage ``RK443'' scheme.\citep{AscherRS1997} \replace{}{Interested readers are referred to the Dedalus online tutorial for how the incompressible constraint is implemented. Because the pressure is only determined up to an additive constant, we fix the gauge by requiring the averaged pressure to equal to unity at the lower boundary.}

\subsection{Ideal MHD Equilibria\protect\label{subsec:Ideal-MHD-Equilibria} }

\begin{figure}
\includegraphics[width=1\columnwidth]{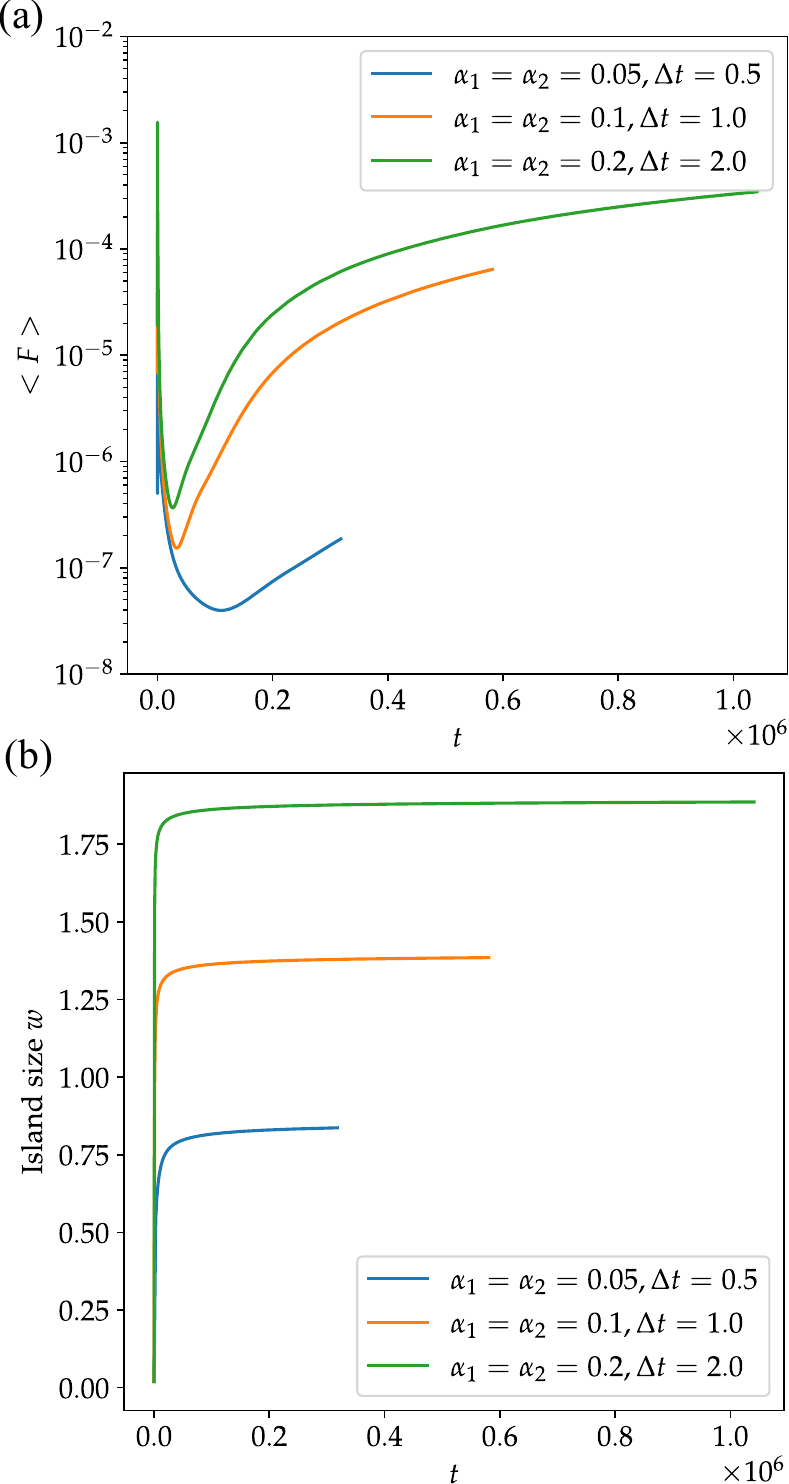}

\caption{The averaged residual force (a) and the magnetic island size (b) versus
time for runs with different Voigt coefficients and time steps. \protect\label{fig:Tearing-ideal-convergence}}
\end{figure}
\begin{figure}
\includegraphics[width=1\columnwidth]{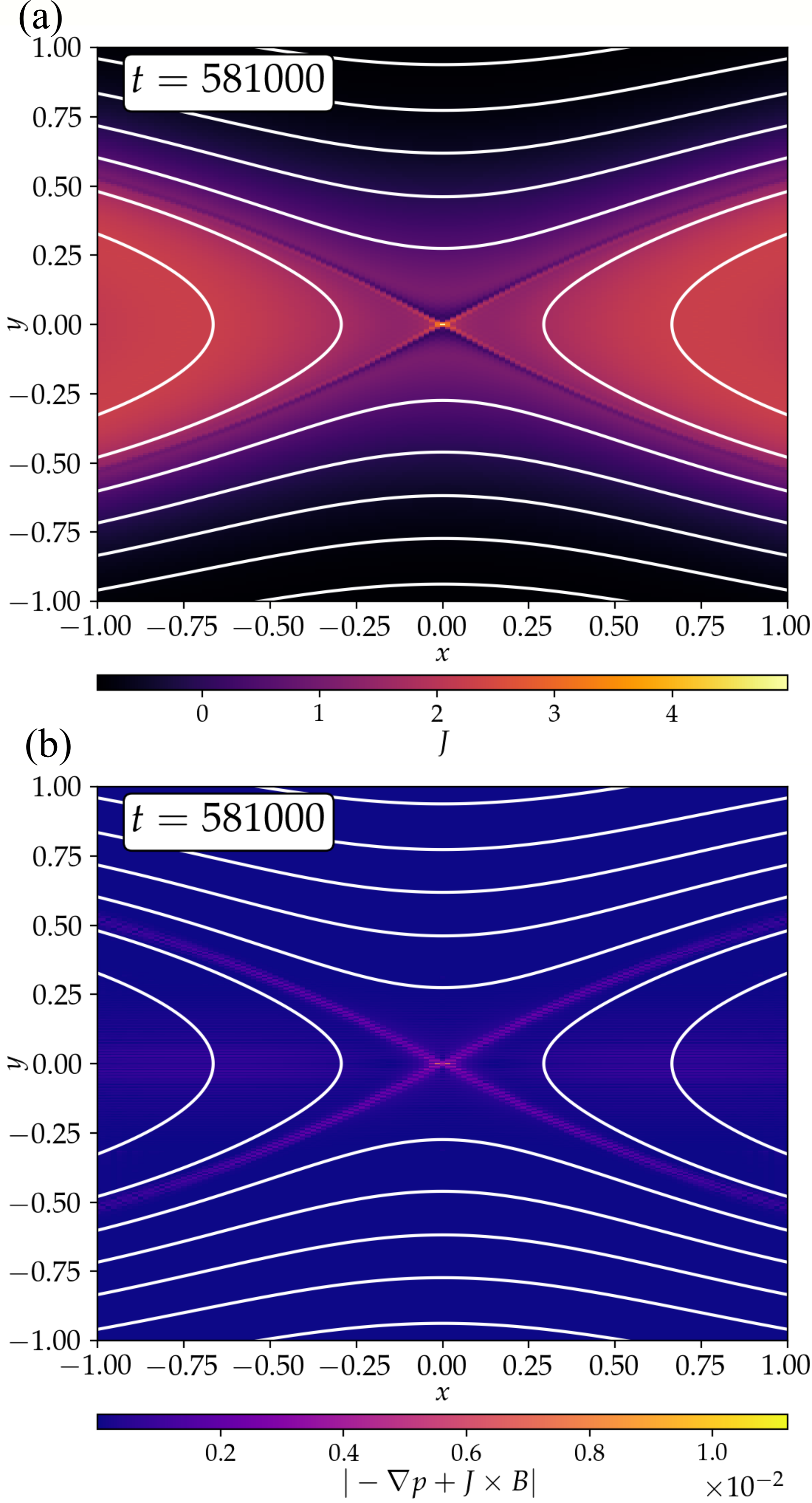}

\caption{(a) A zoom-in view of the current density distribution near the X-point
for the run with $\alpha_{1}=\alpha_{2}=0.1$ at a late time. \replace{}{(b) The residual force is concentrated along the separatrix.} \protect\label{fig:Tearing_current}}
\end{figure}

Now we apply the 2D Voigt-MHD code to some test problems. The first
problem is to find an ideal MHD equilibrium, starting from a current
sheet that is unstable to the tearing instability. We start from a
one-dimensional equilibrium defined by\cite{LoizuHHBKQ2020}
\begin{equation}
\psi_{0}=a\left(\frac{1}{\cosh^{2}\left(y\right)}-1\right),\label{eq:tearing_IC}
\end{equation}
where the coefficient $a=3\sqrt{3}/4$. The corresponding current
density is 
\begin{equation}
J_{0}=a\left(\frac{2}{\cosh^{4}\left(y\right)}-\frac{4\tanh^{2}\left(y\right)}{\cosh^{2}\left(y\right)}\right).\label{eq:J0}
\end{equation}
We perturb the field by first adding $-\epsilon\left(L_x/2\pi\right)\cos\left(2\pi x/L_x\right)\cos\left(\pi y/L_y\right)$ to $\psi_{0}$, then multiplying the resulting flux function by a factor of $1+\epsilon\cos\left(2\pi x/L_{x}\right)$, where $\epsilon$ is a small parameter. This perturbation seeds a narrow island in the field. 

In this set of runs, the resistivity $\eta$ and the external electric
field $E_{\text{ext}}$ are set to zero. The Voigt term in the induction
equation facilitates reconnection, allowing the island to grow. We
set $L_{x}=4$, $L_{y}=2\pi,$ and $\epsilon=10^{-4}$. The resolution
is $N_{x}=N_{y}=256$. The viscosity is set to $\nu=10^{-2}.$ We
vary the Voigt coefficients for the following values $\alpha_{1}=\alpha_{2}=0.05,$
$0.1$, and $0.2$; the corresponding time steps are $\Delta t=0.5$,
$1$, and $2$.

We test the convergence of the numerical solution toward the infinite-time
asymptotic equilibrium by evaluating the averaged residual force over
the whole volume
\begin{equation}
\left\langle F\right\rangle =\left\langle \left|-\nabla p+\boldsymbol{J}\times\boldsymbol{B}\right|\right\rangle .\label{eq:residual_f}
\end{equation}
Figure \ref{fig:Tearing-ideal-convergence}(a) shows the averaged
residual force versus time for various runs. For all these runs, the
averaged residual force initially tends to decrease rapidly, but eventually this tendency
is arrested. Worse, the averaged residual force starts to increase
gradually. As such, the numerical solutions of these runs do not appear
to satisfactorily converge to asymptotic equilibria. Figure \ref{fig:Tearing-ideal-convergence}(b)
plots the magnetic island size versus time for various runs, showing
that the island size is still slowly growing at later times.

The residual force distribution [Fig.~\ref{fig:Tearing_current}(b)] reveals that the unbalanced
force is mostly localized along the separatrix near the X-point, where
the current density $J$ becomes spiky [Fig.~\ref{fig:Tearing_current}(a)].
Inspecting the current density profile also suggests that this numerical
calculation does not have sufficient resolution to resolve the spiky
current density. Moreover, because the force balance condition $-\nabla p+\boldsymbol{J}\times\boldsymbol{B}=0$
in 2D implies $\boldsymbol{B}\cdot\nabla J=0$, the spiky $J$ near
the X-point should spread out exactly along the separatrix such that
$J=J\left(\psi\right)$. This condition is clearly not satisfied in
Figure \ref{fig:Tearing_current}(a). To resolve this thin current structure,
we must have high resolution not only near the X-point, but also along
the separatrix. However, even if we have sufficient numerical resolution,
the spreading of current density along the separatrix would likely take a long time. 

Although our numerical solutions have not achieved desirable convergence
and the magnetic island sizes are still slowly growing, it is clear
that the final equilibrium depends on the Voigt coefficient $\alpha_{2}$.
This fact can be inferred from Eq.~(\ref{eq:induction2d}) as follows.
Because $\boldsymbol{u}=0$ at the X-point and the O-point, the quantity
$\psi+\alpha_{2}J$ is time-independent at these two points. Neglecting
the small initial seeded island, the reconnected magnetic flux is
given by $\psi_{\text{O}}-\psi_{\text{X}}=\alpha_{2}\left(J_{\text{X}}-J_{\text{O}}\right)$,
where the subscripts X and O specify the values evaluated at the X-point
and the O-point, respectively. If the asymptotic solution is independent
of $\alpha_{2}$, then both $\psi_{\text{O}}-\psi_{\text{X}}$ and
$J_{\text{X}}-J_{\text{O}}$ are independent of $\alpha_{2}$, leading
to a contradiction.

\subsection{Resistive MHD Equilibria with External Electric Field \protect\label{subsec:Resistive-MHD-Equilibria}}

\begin{figure}
\includegraphics[width=1\columnwidth]{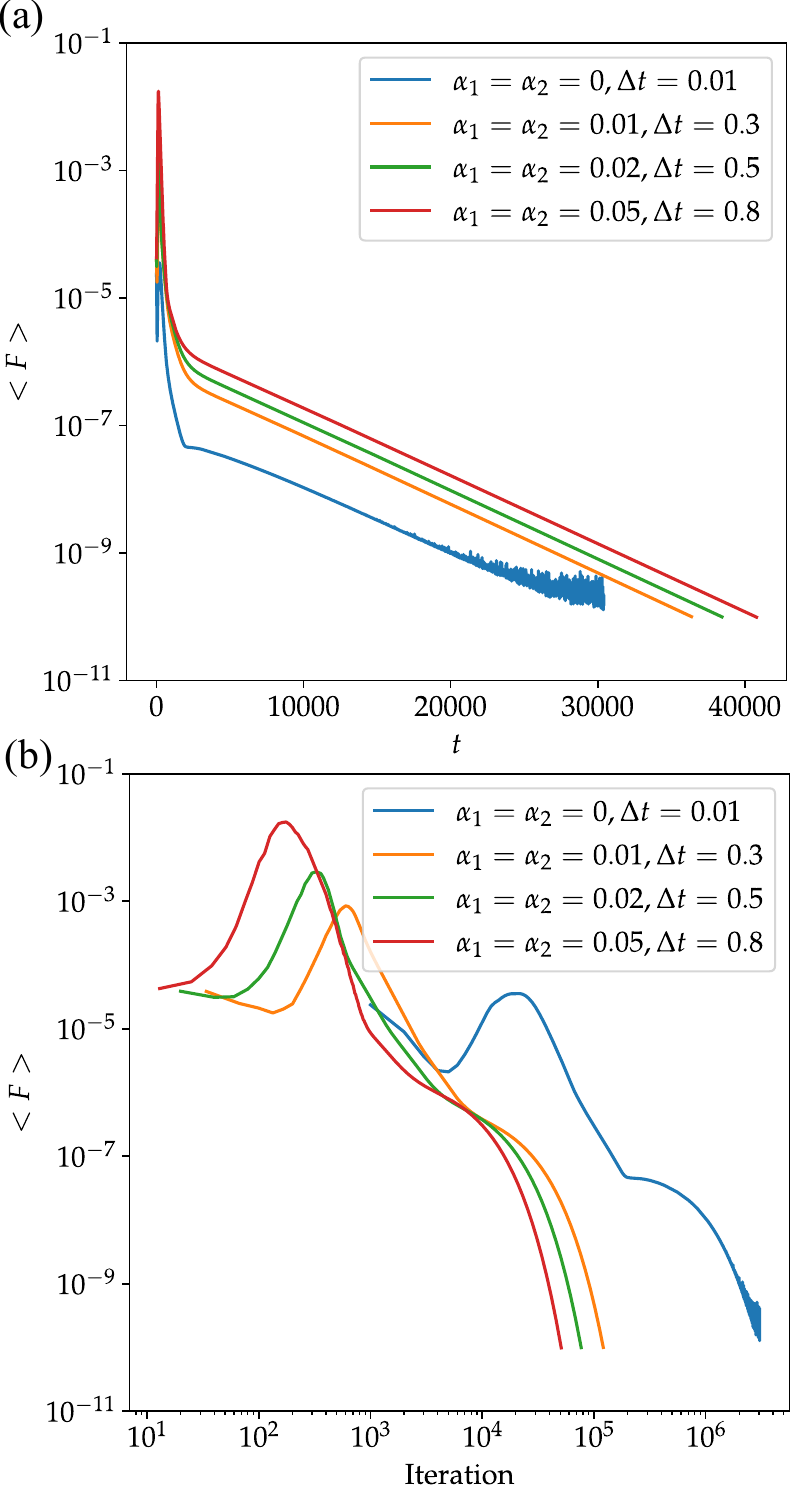}

\caption{The averaged residual force versus time (a) and the number of iterations
(b) for resistive tearing runs. \protect\label{fig:Tearing-res-convergence}}
\end{figure}
\begin{figure}
\includegraphics[width=1\columnwidth]{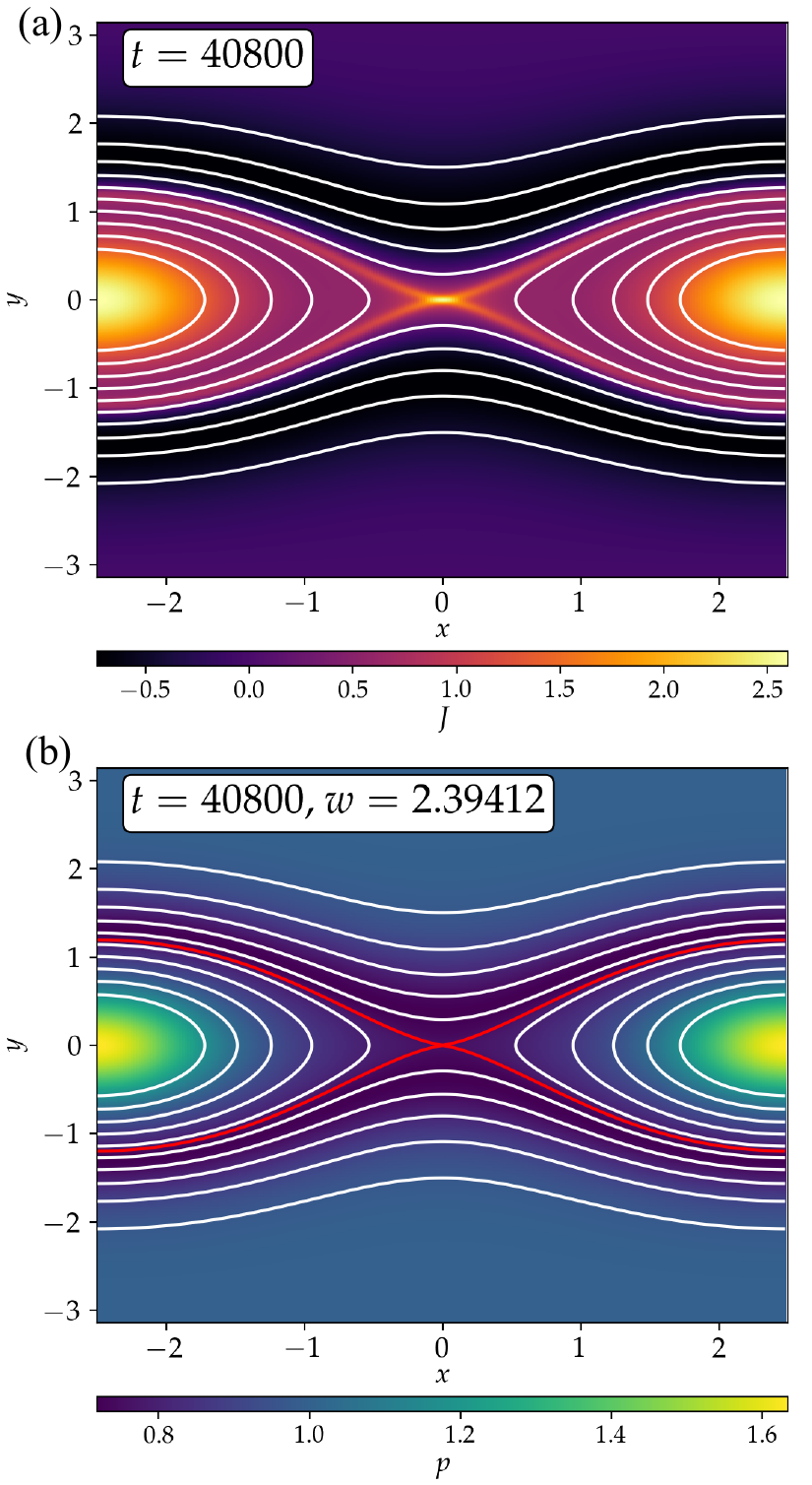}

\caption{The current density (a) the pressure (b) distribution of converged
equilibrium of the resistive tearing problem, from the calculation
of $\alpha_{1}=\alpha_{2}=0.05$. White contours are samples of {magnetic surfaces}, and the red contour denotes the saturated magnetic island.\protect\label{fig:Tearing_resistive}}
\end{figure}

{Now we continue our test with added resistivity.} \replace{}{We examine two distinct resistive scenarios: the saturation of a spontaneous tearing instability, continuing the setup from Sec.~\ref{subsec:Ideal-MHD-Equilibria}, and the Hahm-Kulsrud-Taylor forced reconnection problem, which represents a stable initial state driven by boundary perturbations.}
To prevent the solution from decaying to a vacuum field, an external
electric field $E_{\text{ext}}=-\eta J_{0}$ is applied.

With a finite resistivity, the plasma flow in the final equilibrium
no longer vanishes. Roughly speaking, the plasma flow is proportional
to $\eta$. Because of the non-vanishing flow, the $\boldsymbol{u}\cdot\nabla\boldsymbol{u}$
term and the viscous term must be incorporated in the calculation
of the averaged residual force:
\begin{equation}
\left\langle F\right\rangle =\left\langle \left|\nabla p+\boldsymbol{J}\times\boldsymbol{B}-\boldsymbol{u}\cdot\nabla\boldsymbol{u}+\nu\nabla^{2}\boldsymbol{u}\right|\right\rangle .\label{eq:ave_F}
\end{equation}

 For the resistive tearing mode runs, we set $\eta=\nu=10^{-3}$. The system size is given by $L_{x}=5$ and $L_{y}=2\pi$. The perturbation coefficient is $\epsilon=10^{-4}$.
The resolution is $N_{x}=N_{y}=256$. We vary the Voigt coefficients for the following values $\alpha_{1}=\alpha_{2}=0,$
$0.01$, $0.02$, and $0.05$; the corresponding time steps are $\Delta t=0.01$,
$0.3,$ $0.5$, and $0.8$. Note that because of the flow, which ultimately
limits the {numerical stability} condition, further increasing the Voigt coefficients
does not allow significantly larger time-steps. 

We run the simulations until the averaged residual force $\left\langle F\right\rangle \le10^{-10}$.
Figure \ref{fig:Tearing-res-convergence}(a) shows the averaged residual
force versus time for different Voigt coefficients. The results indicate
that cases with larger Voigt coefficients take longer to converge
with respect to the time variable. In practice, what really matters
is the number of iterations it takes to obtain the converged solution.
As can be seen in Figure \ref{fig:Tearing-res-convergence} (b), cases
with larger Voigt coefficients take fewer iterations to converge,
because of the larger time-steps. Without Voigt regularization, it
takes approximately three million iterations to achieve $\left\langle F\right\rangle <10^{-10}$,
whereas with Voigt coefficients $\alpha_{1}=\alpha_{2}=0.05$, it
takes approximately $51,000$ iterations for the same accuracy. That
is approximately a factor of 60 speed-up.

The converged solutions are practically independent of the Voigt coefficients
and time-steps. \replace{}{To quantify the difference between solutions, we measure the relative error by using the $\alpha_1=\alpha_2=0$ solution as the reference. For the cases $\alpha_1=\alpha_2=0.01$, $0.02$, and $0.05$, the relative errors in the magnetic field, $\left<\delta B\right>/\left<B\right>$, are $3.80\times 10^{-5}$, $4.26\times 10^{-5}$, and $4.56 \times 10^{-5}$, respectively.}

Figure \ref{fig:Tearing_resistive} shows the current
density and pressure distributions of the converged solution, obtained
from the run with $\alpha_{1}=\alpha_{2}=0.05$, overlaid with field
lines (white contours) and the outline of the magnetic island (red
contours). The saturated island size $w\approx2.3941$. Note that
the current density near the X-point is smoother than the ideal case
shown in Figure \ref{fig:Tearing_current}. Importantly, the presence
of flow in the resistive case means the equilibrium no longer satisfies
$\boldsymbol{B}\cdot\nabla J=0$, a condition that holds strictly
in 2D ideal MHS equilibria. Therefore, the current density peak at the X-point does
not exactly spread out along the separatrix. 

\begin{figure}
\includegraphics[width=1\columnwidth]{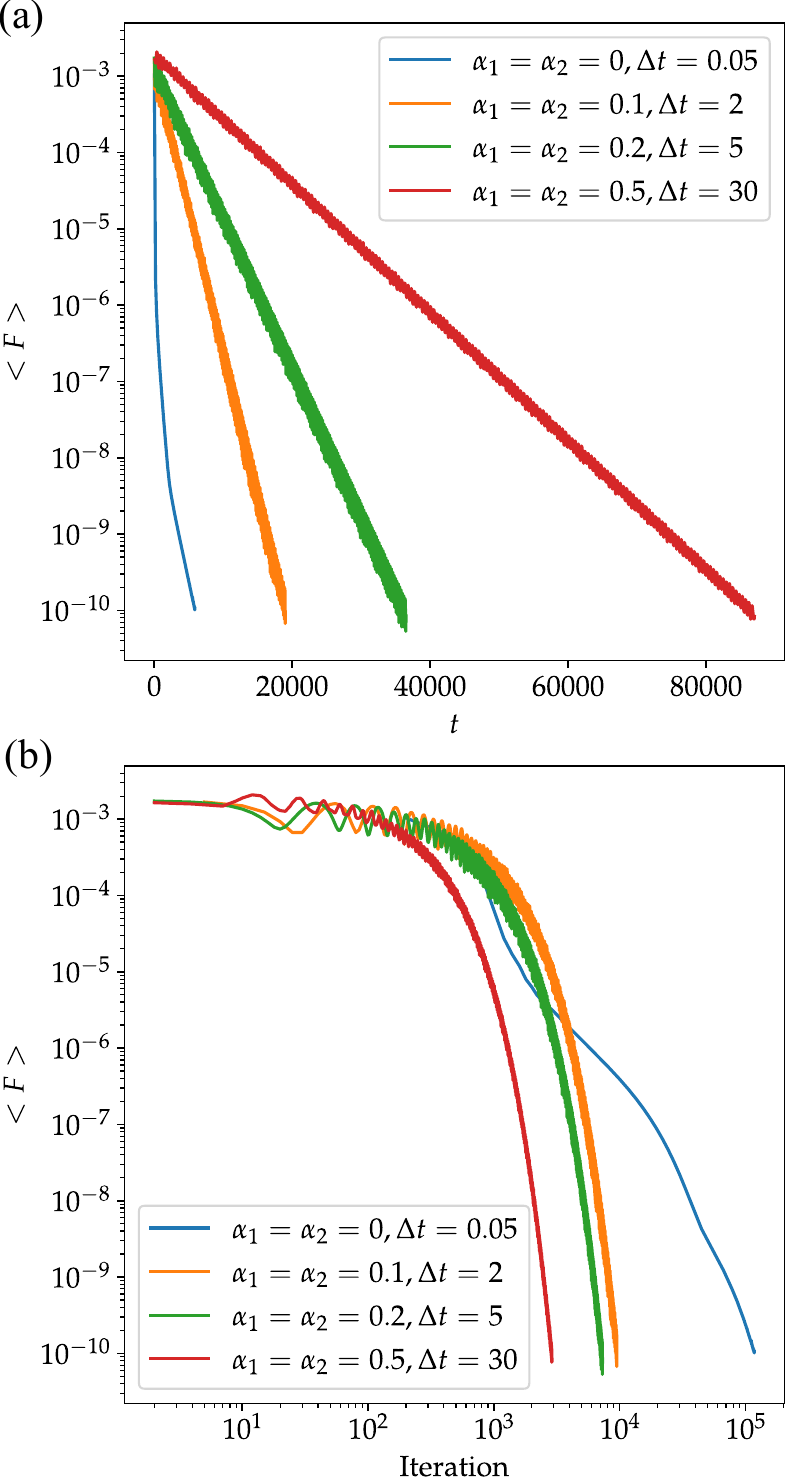}

\caption{The averaged residual force versus time (a) and the number of iterations
(b). \protect\label{fig:HKT-convergence}}
\end{figure}
\begin{figure}
\includegraphics[width=1\columnwidth]{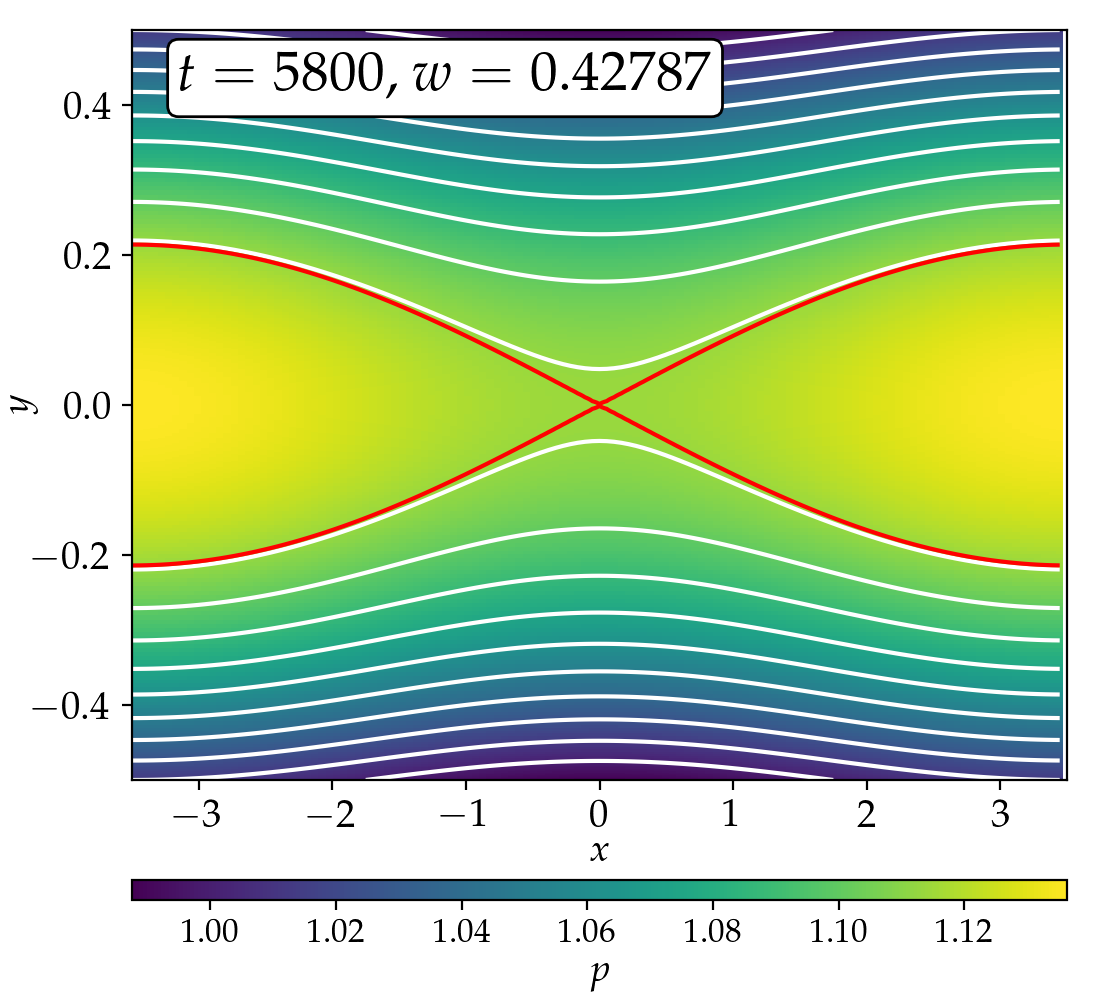}

\caption{The pressure distribution of converged equilibrium of the HKT problem,
from the calculation of $\alpha_{1}=\alpha_{2}=0$. White contours
are samples of {magnetic surfaces}, and the red contour denotes the saturated
magnetic island.\protect\label{fig:HKT-pressure}}
\end{figure}

The next test problem is the Hahm--Kulsrud--Taylor forced reconnection
problem.\citep{HahmK1985} The initial condition is 
\begin{equation}
\psi=\frac{1}{2}y^{2}\left(1+\epsilon\cos\left(2\pi x/L_{x}\right)\right),\label{eq:HKT_initial}
\end{equation}
corresponding to a uniformly sheared magnetic field $\boldsymbol{B}=y\boldsymbol{\hat{x}}$
with an added perturbation. We set $L_{x}=7$, $L_{y}=1,$ and $\epsilon=0.1$.
The resolution is $N_{x}=N_{y}=128$. The viscosity and resistivity
are set to $\eta=\nu=10^{-4}.$ We vary the Voigt coefficients for
the following values $\alpha_{1}=\alpha_{2}=0,$ $0.1$, $0.2$, and
$0.5$. The corresponding time-steps are $\Delta t=0.05$, $2$, $5$,
and $30$. In this example, the plasma velocity asymptotically goes
to zero. Therefore, the stability condition is only limited by waves, and
we can use much larger time-steps with the Voigt regularization.

We again run the simulations until the averaged residual force $\left\langle F\right\rangle \le10^{-10}$.
Figure \ref{fig:HKT-convergence}(a) shows the averaged residual force
versus time for different Voigt coefficients. Again, cases with larger
Voigt coefficients take longer to converge with respect to the time
variable. On the other hand, as shown in Figure \ref{fig:HKT-convergence}
(b), cases with larger Voigt coefficients take fewer iterations to
converge, because of the larger time-steps. Without Voigt regularization,
it takes approximately $120,000$ iterations to achieve $\left\langle F\right\rangle <10^{-10}$,
whereas with Voigt coefficients $\alpha_{1}=\alpha_{2}=0.5$, it only
takes approximately $2,900$ iterations for the same accuracy, corresponding
to a factor of 40 speed-up.

The converged solutions are again \replace{}{nearly} independent of the Voigt coefficients
and time steps. \replace{}{Using the $\alpha_1=\alpha_2=0$ solution as the reference, the relative errors in the magnetic field for the cases $\alpha_1=\alpha_2=0.1$, $0.2$, and $0.5$ are $1.39\times 10^{-5}$, $1.43\times 10^{-5}$, and $1.43 \times 10^{-5}$, respectively.}  Figure \ref{fig:HKT-pressure} shows the pressure
profile of the converged solution, obtained from the run without Voigt
regularization, overlaid with {magnetic surfaces} (white contours) and the
outline of the magnetic island (red contours). The saturated island
size $w\approx0.427872$. 

\replace{}{A curious feature of the time evolultion of the residual force shown in Fig.~\ref{fig:HKT-convergence} is the oscillatory behavior when the Voigt regularization is turned on. This behavior is due to the fact that the Voigt term in the induction equation makes reconnection so fast that the island size overshoots before it relaxes to the final equilibrium through a sequence of oscillations. The animation in the supplementary material illustrates this behavior for the case $\alpha_1=\alpha_2=0.2$. }

\section{Conclusions and Future Perspectives\protect\label{sec:Conclusions-and-Future}}

In summary, this study represents the first investigation of
Voigt regularization for obtaining MHD equilibria, demonstrating its
potential for accelerating the convergence to solutions in resistive
MHD problems while also highlighting challenges in applying the method
to ideal MHD systems.

Despite the theorem of Constantin and Pasqualotto guaranteeing regular time-asymptotic solutions, our calculations of ideal MHD equilibria encounter difficulties in achieving full convergence, primarily due to insufficient resolution of the spiky current density near the X-point. Moreover, the strict requirement of current density being constant along magnetic field lines in 2D ideal MHS equilibria, coupled with the spiky current density at the X-point, may also require an impractically long time to relax to an equilibrium.

In contrast, our numerical experiments demonstrate that Voigt regularization
can be a powerful technique to obtain resistive MHD equilibria with
an external electric field. Because the Voigt terms slow down MHD
waves, thereby allowing larger time-steps, the convergence to equilibrium
solutions is significantly accelerated without affecting the final
equilibrium state.

Our resistive MHD test problems serve as prototypes for more general
problems incorporating transport effects and source terms. Examples include problems of slowly diffusing resistive equilibrium with particle sources\citep{KruskalK1958,GradH1970,HuangH2004a}
or problems with thermal conduction and heat sources. In practice, equilibria
with sources and transports are physically more relevant
than ideal MHD equilibria. Plasma flow that naturally arises in this type of equilibria may also mitigate some of the problems caused by the restrictions of ideal MHD equilibria, such as the strict enforcement of $\boldsymbol{B}\cdot\nabla p=0$ and $\boldsymbol{B}\cdot\nabla J=0$ (in 2D). 

Many questions remain open for future investigations. An immediate
next step is to increase the resolution in the ideal MHD test problem
in Section \ref{subsec:Ideal-MHD-Equilibria} to see if converged solutions can be obtained, perhaps using adaptive mesh refinemnt techniques.\citep{bhattacharjee2005} Three-dimensional generalizations of the
problem that allow formation of chaotic field line region should be attempted, as that could help alleviate some difficulties in Section
\ref{subsec:Ideal-MHD-Equilibria} due to 2D symmetry. Another question
we have not explored deeply is how to optimally choose Voigt coefficients
and time-steps to achieve fast convergence with a small number of iterations. Techniques such as pseudotransient continuation may further improve the convergence speed.\citep{CoffeyKK2003} Finally, a generalization of Voigt regularization to compressible
MHD, incorporating transports and various sources, should be explored.
This could lead to development of more efficient and physically relevant
MHD equilibrium solvers, further contribute to the design and optimization
of future fusion devices.

\section*{Supplementary Material}

See the supplementary material for an animation showing the time evolution of the pressure and the island size for the resistive HKT case with $\alpha_1=\alpha_2=0.2$.

\section*{Data Availability}
The data supporting the findings of this study are openly
available at the following URL/DOI: \url{https://doi.org/10.5281/zenodo.15347128}.

\begin{acknowledgments}
We dedicate this article to Prof. Robert (Bob) Dewar, who suggested
this approach and had many insightful discussions with the authors before
his untimely passing. \replace{}{We thank the referees for suggestions that greatly improved the presentation of this article.} This research was supported by the Simons Foundation/SFARI
(grant No. 560651, A.B.) and the Department of Energy SciDAC HifiStell grant (DE-SC0024548). Part of the computations were performed
on facilities at the National Energy Research Scientific Computing
Center. 
\end{acknowledgments}

\bibliographystyle{apsrev4-1}
\bibliography{Voigt}

\end{document}